\documentclass[twocolumn,amsmath,amssymb,prd,showpacs]{revtex4}

\usepackage{graphicx}

\def\al{\alpha}
\def\be{\beta}
\def\ga{\gamma}
\def\de{\delta}

\def\ve{\varepsilon}

\def\et{\eta}
\def\th{\theta}

\def\la{\lambda}

\def\rh{\rho}

\def\si{\sigma}

\def\ta{\tau}

\def\ph{\phi}

\def\om{\omega}
\def\Ga{\Gamma}

\def\Th{\Theta}
\def\La{\Lambda}
\def\Si{\Sigma}

\def\Om{\Omega}

\def\cl{{\cal L}}

\def\fr#1#2{{{#1} \over {#2}}}
\def\half{{\textstyle{1\over 2}}}

\def\lsim{\mathrel{\rlap{\lower3pt\hbox{$\sim$}}
    \raise2pt\hbox{$<$}}}
\def\gsim{\mathrel{\rlap{\lower3pt\hbox{$\sim$}}
    \raise2pt\hbox{$>$}}}
\def\sqr#1#2{{\vcenter{\vbox{\hrule height.#2pt
         \hbox{\vrule width.#2pt height#1pt \kern#1pt
         \vrule width.#2pt}
         \hrule height.#2pt}}}}

\def\prt{\partial}
\def\lrpartial{\raise 1pt\hbox{$\stackrel\leftrightarrow\partial$}}

\def\Re{\hbox{Re}\,}
\def\Im{\hbox{Im}\,}

\def\etal{{\it et al.}}

\newcommand{\beq}{\begin{equation}}
\newcommand{\eeq}{\end{equation}}
\newcommand{\bea}{\begin{eqnarray}}
\newcommand{\eea}{\end{eqnarray}}
\newcommand{\bit}{\begin{itemize}}
\newcommand{\eit}{\end{itemize}}
\newcommand{\rf}[1]{(\ref{#1})}

\begin{document}

\title{The \v{C}erenkov effect in Lorentz-violating vacua}

\author{Ralf Lehnert}

\author{Robertus Potting}%

\affiliation{CENTRA, Departamento de F\'{\i}sica, 
Universidade do Algarve, 
8000-117 Faro, Portugal}

\date{August 25, 2004}

\begin{abstract}
The emission of electromagnetic radiation
by charges
moving uniformly in a Lorentz-violating vacuum
is studied.
The analysis is performed
within the classical Maxwell--Chern--Simons limit
of the Standard-Model Extension (SME)
and confirms the possibility of a \v{C}erenkov-type effect.
In this context,
various properties of \v{C}erenkov radiation
including the rate, polarization, and propagation features,
are discussed,
and the back-reaction on the charge
is investigated.
An interpretation
of this effect
supplementing the conventional one
is given.
The emerging physical picture
leads to a universal methodology
for studying the \v{C}erenkov effect
in more general situations.
\end{abstract}

\pacs{41.60.Bq, 11.30.Cp, 11.30.Er, 13.85.Tp}

\maketitle

\section{Introduction}

Recent years have witnessed 
substantial progress in quantum-gravity phenomenology: 
minute Lorentz and CPT violations 
have been identified 
as promising candidate signatures for Planck-scale physics \cite{cpt01}. 
At presently attainable energies, 
such signatures are described 
by an effective-field-theory framework 
called the Standard-Model Extension (SME) \cite{ck,grav,kl01}. 
The violation parameters in the SME 
can arise in various underlying contexts, 
such as strings \cite{kps}, 
spacetime-foam approaches \cite{lqg,kp03,klink}, 
noncommutative geometry \cite{ncft}, 
varying scalars \cite{vc,aclm}, 
random-dynamics models \cite{rd}, 
multiverses \cite{mv}, 
and brane-world scenarios \cite{bws}. 
The flat-spacetime limit of the SME 
has provided the basis for 
numerous analyses of Lorentz breaking 
including ones involving
mesons \cite{hadronexpt,kpo,hadronth,ak}, 
baryons \cite{ccexpt,spaceexpt,cane}, 
electrons \cite{eexpt,eexpt2,eexpt3}, 
photons \cite{cfj,photonexpt,photonth,cavexpt,km},
muons \cite{muons},
and the Higgs sector \cite{higgs};
neutrino-oscillation experiments 
offer the potential for discovery 
\cite{ck,neutrinos,nulong}. 

A Lorentz-violating vacuum 
acts in many respects 
like a nontrivial medium. 
For example, 
one expects the electrodynamics limit of the SME
to possess features 
similar to those of ordinary electrodynamics 
in macroscopic media \cite{ck}. 
Indeed, 
changes in the propagation of electromagnetic waves, 
such as modified group velocities and birefringence, 
have been predicted  
and used 
to place tight bounds on Lorentz violation \cite{cfj,km}.
Another conventional feature in macroscopic media, 
which is associated with fast charges, 
is the emission of \v{C}erenkov light 
\cite{cerh,cerexp,cerclass,cerquant}.
A similar mechanism, 
radiation of photons from charged particles
in certain Lorentz-breaking vacua, 
has been suggested in the literature \cite{vcr}. 
Because of the  close analogy 
to the conventional \v{C}erenkov case, 
this mechanism is sometimes called 
the ``vacuum \v{C}erenkov effect.''
This idea is widely employed
in cosmic-ray analyses
of Lorentz violation \cite{vcrtests}. 
The first complete theoretical discussion 
of the vacuum \v{C}erenkov effect---including 
a general methodology 
for extracting radiation rates 
in classical situations---has 
been performed in Ref.\ \cite{lp04}. 
This methodology has subsequently also been employed
in a non-electromagnetic context \cite{aclt04}.

The present work extends our previous analysis \cite{lp04}:
we give a more detailed derivation of the results  
and investigate additional important aspects 
of vacuum \v{C}erenkov radiation. 
More specifically, 
we provide an intuitive physical picture for the effect,
discuss the polarization and propagation properties 
of the emitted light, 
and study the back-reaction on the charge.  
Our analysis is performed 
within the classical Maxwell--Chern--Simons limit
of the SME's electrodynamics sector,  
but we expect our methodology and results
to be applicable in more general situations as well.

A more refined understanding 
of the \v{C}erenkov effect in conventional physics 
provides an additional motivation for our study. 
Recent observations at CERN 
involving high-energy lead ions \cite{cern} 
and experiments in exotic condensed-matter systems \cite{cms} 
have revived the interest in the subject \cite{certheo}. 
In particular, 
these experimental and theoretical investigations
have found evidence 
for unconventional kinematical radiation conditions, 
backward photon emission, 
and backward-pointing radiation cones
in such contexts. 
Since our methodology 
differs from the conventional one 
in several respects, 
it yields additional insight
into the ordinary \v{C}erenkov effect as well. 
For example, 
the presence of a fully relativistic Lagrangian, 
which incorporates dispersion, 
makes it feasible to work in the charge's rest frame 
simplifying the calculation.
In particular,
the exact emission rate for point particles
carrying an electromagnetic charge
and a magnetic moment
can be determined 
without the explicit field solutions. 

The outline of this paper is as follows.
Section \ref{basics} reviews some basics 
of the Maxwell--Chern--Simons model.
In Sec.\ \ref{concept}, 
we set up the problem 
in a model-independent way 
and extract the general condition 
for the emission of \v{C}erenkov light. 
The concrete calculation of the radiation rate 
in the Maxwell--Chern--Simons model
is performed in Sec.\ \ref{rate}.
Section \ref{back} discusses the back-reaction on the charge. 
In Sec.\ \ref{phase}, 
a complementary, 
purely kinematical approach for estimating radiation rates 
is presented. 
We comment briefly on experimental implications 
in Sec.\ \ref{exp}. 
The conclusions are contained in Sec.\ \ref{conc}. 
Appendix \ref{dr} provides supplementary material 
about the plane-wave dispersion relation. 
The wave polarizations are briefly discussed 
in Appendix \ref{pws}. 
In Appendix \ref{md}, 
we determine the radiation rate 
for a charged magnetic dipole.

\section{Basics}
\label{basics}

The renormalizable gauge-invariant photon sector of the SME 
contains a CPT-odd and a CPT-even operator 
parametrized by $(k_{AF})^{\mu}$ and $(k_F)^{\mu\nu\rh\si}$, respectively. 
Many components of these parameters 
are strongly constrained 
by astrophysical spectropolarimetry \cite{photonexpt}. 
However, 
further investigations remain to be of great interest both 
for a better understanding of massless Lorentz-violating fields 
and for the potential of complementary tighter bounds. 
In the present work, 
we consider the Chern--Simons-type $(k_{AF})^{\mu}$ modification.
The Maxwell--Chern--Simons model
resulting from such a modification 
has been studied extensively in the literature \cite{cfj,klink,mcsclass,mcsquant}.
The $(k_{AF})^{\mu}$ parameter has mass dimensions,
which leads to the more interesting case of nontrivial dispersion.
Note also 
that many characteristics of \v{C}erenkov radiation, 
such as rates and energy fluxes,
require at least one dimensionful model parameter. 
With only a dimensionless $(k_F)^{\mu\nu\rh\si}$,
the usual approach to the \v{C}erenkov effect 
involving an external nondynamical point source
might be problematic for obtaining a finite rate.

In natural units 
$c\hspace{-1pt} =\hspace{-1pt} \hbar\hspace{-1pt} =\hspace{-1pt} 1$,
the Maxwell--Chern--Simons Lagrangian
in the presence of external sources $j^{\mu}$
is
\beq
\cl_{\rm MCS} = 
-\fr{1}{4} F_{\mu\nu}F^{\mu\nu}
+(k_{AF})_{\mu}A_{\nu}\tilde{F}^{\mu\nu}
-A_{\mu}j^{\mu}, 
\label{lagr}
\eeq
where $F_{\mu\nu}=\prt_{\mu}A_{\nu}-\prt_{\nu}A_{\mu}$ 
denotes the conventional electromagnetic field-strength tensor 
and $\tilde{F}^{\mu\nu}=\half\ve^{\mu\nu\rh\si}F_{\rh\si}$ its dual, 
as usual. 
Since we take $(k_{F})^{\mu\nu\rh\si}=0$, 
we can omit the subscript $AF$ 
of the Lorentz- and CPT-violating $(k_{AF})^{\mu}$ parameter 
and set $(k_{AF})^{\mu}\equiv k^{\mu}=(k^0,\vec{k})$. 
In this work, 
$k^{\mu}$ must \textit{not} be confused 
with the traditional notation of Fourier momenta.
A nondynamical fixed $k^{\mu}$ 
determines a special direction in spacetime. 
For example, 
certain features of plane waves propagating along $\vec{k}$ 
might differ from those of waves perpendicular to $\vec{k}$. 
Thus, particle Lorentz symmetry is violated \cite{ck,rl03}. 
However, 
note that the Lagrangian \rf{lagr} transforms as a scalar
under rotations and Lorentz boosts of the reference frame. 
This coordinate independence remains a fundamental principle 
regardless of particle Lorentz breaking. 
It guarantees 
that the physics is left unaffected 
by an observer's choice of coordinates 
and is therefore also called observer Lorentz symmetry \cite{ck,rl03}. 
This principle is essential for our discussion in Sec.\ \ref{concept}.

The Lagrangian \rf{lagr} yields the following
equations of motion for the potentials $A^{\mu}=(A^0,\vec{A})$: 
\beq
\left(\Box \et^{\mu\nu}-\prt^{\mu} \prt^{\nu}
-2\ve^{\mu\nu\rh\si}k_{\rh}\prt_{\si}\right)A_{\nu}
=j^{\mu} .
\label{oddeom}
\eeq
As in conventional electrodynamics, 
current conservation $\prt_{\mu}j^{\mu}=0$ 
emerges as a compatibility requirement. 
For completeness, 
we also give the modified Coulomb 
and Amp\`ere laws contained in Eq.\ \rf{oddeom}: 
\bea
\vec{\nabla}\!\cdot\!\vec{E}-2\vec{k}\!\hspace{0.8pt}\cdot\!\vec{B} & = & \rh ,\nonumber\\
-\dot{\hspace{-1pt}\vec{E}}+\vec{\nabla}\!\times\!\vec{B}
-2k_0\vec{B}+2\vec{k}\!\times\!\vec{E} & = & 
\hspace{1.5pt}\vec{\hspace{-1.5pt}\textit{\j}}\hspace{1pt} .
\label{oddmax}
\eea
The homogeneous Maxwell equations 
remain unchanged 
because the field--potential relationship is the usual one.
The potential $A^0$ is nondynamical, 
and gauge symmetry eliminates another component of $A^{\mu}$, 
so that Eq.\ \rf{oddeom} contains two independent degrees of freedom
paralleling the conventional Maxwell case. 
To fix a gauge,
any of the usual conditions on $A^{\mu}$, 
such as Lorentz or Coulomb gauge,  
can be imposed.
Note, however, 
that there are some differences 
between conventional electrodynamics and the present model 
regarding the equivalence of certain gauge choices.
A more detailed discussion 
of the degrees of freedom and the gauge-fixing process 
is contained in the second paper of Ref.\ \cite{ck}.

The tensor given by
\beq 
\Th^{\mu\nu}=-F^{\mu\al}F^{\nu}{}_{\!\al} 
+\fr{1}{4}\et^{\mu\nu}F^{\al\be}F_{\al\be} 
-k^{\nu}\tilde{F}^{\mu\al}A_{\al} 
\label{emtensor} 
\eeq
is associated with the energy and momentum 
stored in our modified electromagnetic fields. 
Here, 
$\et^{\mu\nu}$ denotes the usual metric with signature $-2$ 
in flat Minkowski space. 
Although the energy--momentum tensor is gauge dependent, 
it changes only by a 3-gradient 
under a gauge transformation
$A^{\mu}\rightarrow A^{\mu}+\prt^{\mu}\La$, 
so that the total 4-momentum obtained by a spatial integration 
remains unaffected. 
More generally, 
the action---and therefore the physics---is gauge invariant 
\cite{fn4}. 
As opposed to the conventional case, 
$\Th^{\mu\nu}$ cannot be symmetrized 
because its antisymmetric part 
is no longer a total derivative. 
It follows from the equations of motion \rf{oddeom} 
that the energy--momentum tensor \rf{emtensor} 
obeys 
\beq
\prt_{\mu}\Th^{\mu\nu}=j_{\mu}F^{\mu\nu}. 
\label{emcurrent} 
\eeq
The presence of sources $j^{\mu}\neq 0$ implies
that $\Th^{\mu\nu}$ is not conserved,
as expected \cite{fn1}.

To find solutions of the equations of motion \rf{oddeom}
one can employ standard Fourier methods.
The modified Maxwell operator 
appearing in parentheses in Eq.\ \rf{oddeom}
is singular, 
as in the conventional case. 
This can be verified in Fourier space,
where the corresponding Minkowski matrix fails to be invertible. 
To circumvent this obstacle, 
we can proceed in Lorentz gauge, 
as usual.
Then,
Eq.\ \rf{oddeom} takes the form  
\beq
\left(-p^2 \et_{\nu\la}
+2i\ve_{\nu\la\rh\si}k^{\rh}p^{\si}\right)\hat{A\:\,}\!\!^{\la}
=\textit{\^{\j}}\hspace{.5pt}_{\nu}
\label{moddeom}
\eeq
in $p^{\mu}$ Fourier space.
Here, 
the caret denotes the four-dimensional Fourier transform, 
and the dependence of $\hat{A\:\,}\!\!_{\nu}$ and $\textit{\^{\j}}\hspace{.5pt}_{\nu}$
on the wave 4-vector $p^{\mu}=(\om,\vec{p}\hspace{1pt})$ is understood.
Contraction of Eq.\ \rf{moddeom} 
with the tensor 
\beq
\hat{G}^{\mu\nu} \equiv -\fr{p^2 \et^{\mu\nu}
+2i\ve^{\mu\nu\rh\si}k_{\rh}p_{\si}
+4k^{\mu}k^{\nu}}{p^4+4p^2k^2-4(p\!\hspace{0.8pt}\cdot\! k)^2}+4\hat{G}^{\mu\nu}_0 ,
\label{oddgreen}
\eeq
where
\beq
\hat{G}^{\mu\nu}_0 \equiv
\fr{(p\!\hspace{0.8pt}\cdot\!k)
(p^{\mu}k^{\nu}+k^{\mu}p^{\nu})
-k^2 p^{\mu}p^{\nu}}{\big[p^4+4p^2k^2-4(p\!\hspace{0.8pt}\cdot\!k)^2\big]p^2},
\label{spurious}
\eeq
yields the Fourier-space solution
$\hat{A\:\,}\!\!^{\mu}=\hat{G}^{\mu\nu} \textit{\^{\j}}\hspace{.5pt}_{\nu}$
of the equations of motion. 
This establishes that $\hat{G}^{\mu\nu}(p^{\mu})$
is a momentum-space Green function for Eq.\ \rf{moddeom}.
Transformation to position space
now gives the general solutions $A^{\mu}(x)$
of Eq.\ \rf{oddeom}:
\beq
A^{\mu}(x)=A^{\mu}_0(x)
+\int\limits_{C_\om}\fr{d^4 p}{(2\pi)^4}\;\hat{G}^{\mu\nu}\textit{\^{\j}}\hspace{.5pt}_{\nu}
\exp (-ip\!\hspace{0.8pt} \cdot \! x),
\label{oddsln}
\eeq
where $x^{\mu}=(t,\vec{r}\hspace{1pt})$ 
denotes the spacetime-position vector 
and $A^{\mu}_0(x)$ satisfies Eq.\ \rf{oddeom} in the absence of sources.
As in the conventional case, 
the freedom in choosing $A^{\mu}_0(x)$ 
and the $\om$-integration contour $C_{\om}$
can be used to satisfy boundary conditions. 

The poles of the integrand in Eq.\ \rf{oddsln}
determine the plane-wave dispersion relation. 
With Def.\ \rf{oddgreen}, we obtain
\beq
p^4+4p^2k^2-4(p\!\hspace{0.8pt}\cdot\! k)^2=0 .
\label{odddisp}
\eeq
This equation yields the wave frequency $\om$
for a given wave 3-vector $\vec{p}$.
It is known 
that this dispersion relation 
admits spacelike wave 4-vectors
for any nontrivial value of $k^{\mu}$ \cite{klink}. 
Such vectors
will turn out 
to be the driving entity 
for \v{C}erenkov radiation.
The roots of the dispersion relation \rf{odddisp} 
are discussed in Appendix \ref{dr}. 
Note that the $\hat{G}^{\mu\nu}_0$ term 
exhibits additional poles at $p^2=0$.
However,
this piece of the Green function
does not contribute to the physical fields: 
upon contraction of $\hat{G}^{\mu\nu}_0$ with $\textit{\^{\j}}\hspace{.5pt}_{\nu}$,
terms with $p\cdot\hspace{-0.8pt}\textit{\^{\j}}\hspace{.5pt}$ vanish due to current conservation.
The remaining term in Eq.\ \rf{spurious},
proportional to $p^{\mu}$, 
amounts to a total derivative 
and is therefore pure gauge. 

The contour plays a pivotal role in our study.
We choose the usual retarded boundary conditions, 
so that $C_{\om}$ passes above all poles on the real-$\om$ axis.
However, 
for timelike $k^{\mu}$
Eq.\ \rf{odddisp}
determines poles 
both in the lower and in the upper half plane.
Thus,
the Green function for timelike $k^{\mu}$ 
will in general be nonzero also in the acausal region $t<0$. 
This is consistent 
with earlier findings of microcausality violations 
in the presence of a timelike $k^{\mu}$ \cite{cfj,klink}. 
We remark 
that the Lorentz-violating modifications 
to the conventional Maxwell operator 
leave unchanged the structure of the highest-derivative terms
and are thus mild enough to maintain hyperbolicity. 
Therefore, 
the more general Fourier--Laplace method
can be employed to define a retarded Green function 
in the timelike-$k^{\mu}$ situation 
at the cost of introducing exponentially growing solutions. 
We disregard this possibility here 
and focus on the spacelike- and lightlike-$k^{\mu}$ cases. 

In the present context, 
it is convenient to implement the definition of the contour 
by shifting the poles at real $\om$ into the lower half plane.
To this end, 
we replace $\om\rightarrow\om+i\ve$ in the denominator
of the integrand in Eq.\ \rf{oddsln}. 
Here, 
$\ve$ is an infinitesimal positive parameter 
that is taken to approach zero after the integration. 
This prescription 
is reminiscent of the Fourier--Laplace approach,
where $\ve$ could in general also be finite. 
To ensure compatibility 
of the Fourier--Laplace transform
with the present Fourier methods 
in the limit $\ve\rightarrow +0$, 
the prescription $\om\rightarrow\om+i\ve$
must be implemented 
in each $p$ of the denominator.

\section{Conceptual considerations} 
\label{concept} 

Observer invariance of the Lagrangian \rf{lagr} implies
that the presence or absence of vacuum \v{C}erenkov radiation 
is independent of the coordinate system.
We can therefore proceed in a rest frame of the charge \cite{fn2}.
In such a frame, 
many conceptual issues become more transparent
and the calculations are simpler. 
For instance,
a condition for radiation is
that at least part of the energy--momentum flux
associated with the fields of the charge 
must escape to spatial infinity $|\vec{r}\hspace{1pt}|=r\to\infty$.
Thus, 
the modified energy--momentum tensor \rf{emtensor} 
must contain pieces 
that do not fall off faster than $r^{-2}$. 
Excluding a logarithmic behavior of the potentials, 
both $A^{\mu}$ and $F^{\mu\nu}$ 
should then exhibit
an asymptotic $r^{-1}$ dependence.

To some extent, 
this parallels the case 
of a local time-dependent 4-current distribution 
in conventional electrodynamics. 
For example, 
the fields of a rotating electric dipole 
behave like $r^{-1}\cos(\om t - |\vec{p}\hspace{1pt}|r+\varphi)$
at large distances.
In this example,
$\om$ corresponds to the rotation frequency,
and $\varphi$ denotes a field-specific phase \cite{Jackson}.
Note 
that the fields oscillate both with time and with distance.
In the present case, 
however, 
we require the 4-current of a particle at rest 
to be stationary 
leading to time-independent fields. 
Radiation can then occur 
in the presence of terms oscillating 
with {\it distance only}. 
Such terms are absent in ordinary vacuum electrodynamics. 
In the present context, 
we must therefore investigate 
whether the Lorentz-violating modification of the fields 
associated with a particle at rest 
can exhibit such an oscillatory behavior. 

As a consequence of the presumed time independence 
in the particle's rest frame, 
the four-dimensional $p^{\mu}$ Fourier transform of the 4-current
is $\textit{\^{\j}}\hspace{1pt}^{\mu}(p^{\mu})=2\pi\de(\om)\textit{\~{\j}}\hspace{1pt}^{\mu}(\vec{p}\hspace{1pt})$,
where the tilde denotes the three-dimensional $\vec{p}$ Fourier transform.
Then, 
the charge's fields 
are generally determined 
by the inverse Fourier transform 
of an expression 
containing $1/D(0,\vec{p}\hspace{1pt})$, 
where $D(\om,\vec{p}\hspace{1pt})=0$
is the plane-wave dispersion relation. 
For example, 
in the present Maxwell--Chern--Simons model
Eq.\ \rf{oddsln} leads (up to homogeneous solutions) 
to the fields  
\beq
A^{\mu}(\vec{r}\hspace{1pt}) = 
\int \fr{d^3 \vec{p}}{(2\pi)^3}\;
\fr{N^{\mu\nu}(\vec{p}\hspace{1pt})\textit{\~{\j}}\hspace{.5pt}_{\nu}(\vec{p}\hspace{1pt})
\exp(i\vec{p}\!\hspace{0.8pt}\cdot\!\vec{r}\hspace{1pt})}
{\vec{p}^{\,4}-4\vec{p}^{\,2}k^2-4(\vec{p}\!\hspace{0.8pt}\cdot\!\vec{k}-i\ve k_0)^2},
\label{Asoln}
\eeq
where we have set 
$N^{\mu\nu}(\vec{p}\hspace{1pt})\equiv\vec{p}^{\,2}\et^{\mu\nu}
-2i\ve^{\mu\nu\rh s}k_{\rh}p_{s}
-4k^{\mu}k^{\nu}$ 
for brevity. 
Latin indices run from 1 to 3. 
Note 
that the previous prescription for the $\om$ integral
automatically defines the integral \rf{Asoln}
in the case of singularities.

Regardless of the presence of poles at real $\vec{p}$,
integrands containing 
$\exp(i\vec{p}\!\hspace{0.8pt}\cdot\!\vec{r}\hspace{1pt})/D(0,\vec{p}\hspace{1pt})$
suggest evaluation of the $|\vec{p}\hspace{1pt}|$ integral
with complex-analysis methods. 
The $|\vec{p}\hspace{1pt}|$ integration then gives 
certain residues of the integrand 
in the complex $|\vec{p}\hspace{1pt}|$ plane,
which typically contain the factor
$\exp (i\vec{p}_0\!\cdot\!\hspace{0.8pt}\vec{r}\hspace{1pt})$ 
in the dispersive case. 
Here, 
$\vec{p}_0$ denotes the location of a pole, 
e.g., 
$D(0,\vec{p}_0)=0$.
Note that for $\Re(\vec{p}_0)\neq \vec{0}$
the residues oscillate with distance, 
whereas $\Im(\vec{p}_0)\neq \vec{0}$ 
implies residues that exponentially decay 
with increasing $r$. 
The fields then typically display 
a qualitatively similar behavior 
because the remaining angular integrations
correspond merely to averaging the residues
over all directions. 
It follows 
that there can be emission of light 
for $\Re(\vec{p}_0)\neq \vec{0}$ 
and $\Im(\vec{p}_0)=\vec{0}$. 
In other words, 
we can expect vacuum \v{C}erenkov radiation only
when there are real $p^{\mu}=(0,\vec{p}\hspace{1pt})$
satisfying the plane-wave dispersion relation 
in the charge's rest frame.

This general result suggests
that the radiated energy is zero 
in the rest frame of the particle.
We will explicitly verify this
in the next section.
The above radiation condition requires 
the presence of spacelike wave 4-vectors $p^2<0$.
Note that such wave vectors
can be associated with negative frequencies
in certain frames.
However, 
this does not necessarily lead to positivity violations and instabilities
because the model could be the low-energy limit 
of an underlying positive-definite theory \cite{vc}. 

We continue by verifying 
that the requirement of spacelike wave 4-vectors $p^{\mu}=(0,\vec{p}\hspace{1pt})$
in the charge's rest frame $\Si$
is consistent with the usual phase-speed condition
in the laboratory frame $\Si'$. 
In the conventional case,
\v{C}erenkov radiation can occur 
when the charge's speed 
equals or exceeds the phase speed 
$c\hspace{0.8pt}'_{ph}(\vec{p}\hspace{1pt}'\hspace{-0.8pt})
=|\om'\hspace{-1.5pt}|/|\vec{p}\hspace{1pt}'\hspace{-1.5pt}|$
of light in the medium 
for some $\vec{p}\hspace{1pt}'\!$.
This clearly requires $c\hspace{0.8pt}'_{ph}
(\vec{p}\hspace{1pt}'\hspace{-0.8pt})<1$
implying $p^2\!\,=p'\!\!\cdot\! p'\!
=\om'^2-\vec{p}\hspace{1pt}'^2<0$,
where we have used coordinate independence.
This first step establishes the need 
for spacelike wave vectors
for conventional \v{C}erenkov radiation.
Next,
we include the condition on the charge's speed
into our discussion.
The above vacuum \v{C}erenkov condition
also requires the specific form $p^{\mu}=(0,\vec{p}\hspace{1pt})$
of the spacelike wave vectors in $\Si$.
If the charge moves with velocity $\vec{\be}'\!$ in $\Si'\!$,
the laboratory-frame components of such a wave 4-vector
are 
$(\vec{\be}'\!\!\cdot\!\vec{p}\hspace{1pt}'\!\!,
\vec{p}\hspace{1pt}')$,
where $\vec{p}\hspace{1pt}'\!=\vec{p}
+(\ga-1)(\vec{p}\hspace{-0.0pt}\cdot\hspace{-0.5pt}\vec{\be}'\hspace{-0.5pt})
\,\vec{\be}'\!/\vec{\be}'^2$
is the wave 3-vector in $\Si'\!$
and $\ga$ the relativistic gamma factor 
corresponding to $|\vec{\be}'\hspace{-1.2pt}|$.
This yields
the conventional condition
$c\hspace{0.8pt}'_{ph}
=|\vec{\be}'\!\!\cdot\!\hspace{0.8pt}\vec{p}\hspace{1pt}'\hspace{-1.2pt}|
/|\vec{p}\hspace{1pt}'\hspace{-1.2pt}|
\le|\vec{\be}'\hspace{-1.2pt}|$.

The physics of \v{C}erenkov radiation 
can now be understood intuitively
as follows. 
In nontrivial vacua 
(e.g., fundamental Lorentz violation 
or conventional macroscopic media) 
the plane-wave dispersion relation
can admit real spacelike wave 4-vectors as solutions.
As opposed to the Lorentz-symmetric case,
the fields of a charge at rest 
can then contain time-independent spatially oscillating exponentials 
in its Fourier decomposition. 
Such waves can carry 4-momentum
to spatial infinity 
implying the possibility of net radiation. 
This conceptual picture also carries over to quantum theory,
where the field of a charge 
can be pictured 
as a cloud of photons of all momenta. 
In conventional QED, 
these photons are {\it virtual}. 
In the present context, 
however, 
{\it real} photons contribute as well. 
Consider, 
for example, 
the one-loop self-energy diagram for a massive fermion. 
One can verify 
that for photons with spacelike $p^{\mu}$ 
there are now loop momenta 
at which both 
the photon and fermion propagator are on-shell. 
Thus, 
the photon cloud of the fermion 
can contain real (spacelike) photons. 
Moreover, 
the diagram can be cut at the two internal lines, 
so that photon emission need not be followed by reabsorption 
and the photon can escape. 

\begin{figure}
\begin{center}
\includegraphics[width=0.8\hsize]{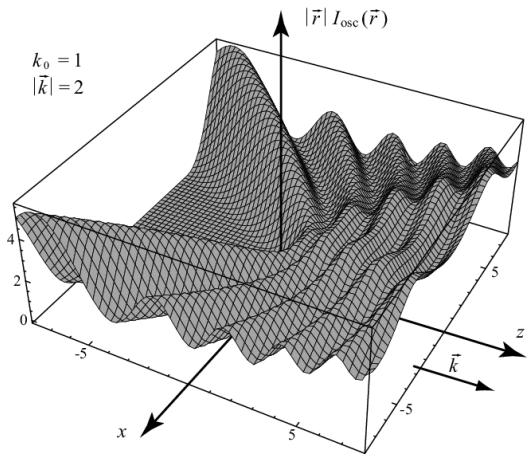}
\end{center}
\caption{General field pattern of a point charge resting at the origin.
The function 
$|\vec{r}\hspace{1pt}|I_{\rm osc}(\vec{r}\hspace{1pt})$ is shown
for $\vec{r}$ in the $xz$ plane 
with $\vec{k}$ along the $z$ direction. 
This function was evaluated
by an analytical  $|\vec{p}\hspace{1pt}|$-type integration followed 
by numerical angular integrations. 
Uninteresting nonoscillatory pieces 
$I_{\rm non}$ have been subtracted for clarity, 
so that only the oscillatory part $I_{\rm osc}\equiv I-I_{\rm non}$ 
contributes to this plot. 
The wave pattern is resemblant 
to that caused by a boat moving in water. 
} 
\label{fig1} 
\end{figure}

The comparison with the wave pattern 
of a boat in calm water
might yield further intuition
(despite differences in the physics involved).
The boat represents the charge  
and surface waves on the water the electromagnetic field.
If the boat rests relative to the water,
waves are absent. 
A moving boat causes
wave patterns emanating from its bow.
An observer on shore 
sees a moving v-shaped wavefront 
with an  opening angle depending on the boat's speed. 
After the wavefront has passed,
the water level on shore oscillates with decaying amplitude.
A passenger on the boat sees
a static wave pattern: 
assuming no turbulence, 
the water surface behind the leading wave front
appears undulated in a time-independent way. 
In our case,
a similar wave pattern forms in the particle's rest frame.
For example, 
consider a point charge $\textit{\~{\j}}\hspace{1pt}^{\mu}=q(1,0)$. 
Equation \rf{Asoln} implies 
$(2\pi)^3A^{\mu}(\vec{r}\hspace{1pt}) 
= qN^{\mu 0}(-i\vec{\nabla})I(\vec{r}\hspace{1pt})$,
where $I(\vec{r}\hspace{1pt})\equiv
\int d^3\vec{p}\;\exp(i\vec{p}\!\hspace{0.8pt}\cdot\!\vec{r}\hspace{1pt})\;
\big[\vec{p}^{\,4}-4\vec{p}^{\,2}k^2
-4(i\ve k_0-\vec{p}\!\hspace{0.8pt}\cdot\!\vec{k})^2\big]^{-1}$
determines the shape of the fields. 
The general behavior of $I(\vec{r}\hspace{1pt})$ follows from Fig.\ \ref{fig1}
and is similar to that of the water waves discussed above.
Note the absence of a shock-wave singularity, 
as expected for nontrivial dispersion. 

We finally remark 
that single-photon emission 
from a massive charged particle 
with conventional dispersion relation $p^2=m^2$ 
requires spacelike photon 4-momenta: 
the initial and final 4-momenta of the charge 
each determine a point on the mass-shell hyperboloid, 
and the photon momentum must connect these two points. 
The geometry of the hyperboloid 
is such that any tangent, 
and therefore any two points, 
determine a spacelike direction. 
Note 
that this is also consistent 
with the above loop-momentum considerations.

\section{Radiation rate}
\label{rate} 

The rate of vacuum \v{C}erenkov radiation 
can in principle be determined 
with the usual philosophy:
extraction of the $r^{-2}$ piece 
of the modified Poynting vector 
and integration over a spherical surface. 
However, 
the integral \rf{Asoln} 
seems to evade a systematic analytical study
(except in special cases),
so that the determination of the asymptotic fields 
is challenging. 
We have therefore employed a method 
for finding the total rate of radiation
that does not require
explicitly the far fields.

Integration of Eq.\ \rf{emcurrent} 
over an arbitrary volume $V$ 
and the divergence theorem imply 
\beq 
\int\limits_{\si} d\si^{l} \, \Th_{l\nu}
= \int\limits_{V}d^3\vec{r} \; j^{\mu}F_{\mu\nu}
- \fr{\prt}{\prt t} \int\limits_{V}d^3\vec{r} \; \Th_{0\nu}\; ,
\label{encons} 
\eeq 
where $\si$ is the boundary of $V$,
and $d\si^{l}$ denotes the associated surface element 
with outward orientation. 
Thus, 
the energy--momentum flux 
$\dot{P}_\nu=\int_{\si} d\si^{l} \, \Th_{l\nu}$
through the surface $\si$ 
originates from the 4-momentum 
generated by the source in the enclosed volume $V$ 
and the decrease of the field's 
4-momentum in $V$, as usual.

We are interested in 4-currents 
describing particles. 
In this case, 
we write $j^{\mu}(x)=J^{\mu}(x)$, 
where $J^{\mu}=(J^0,\vec{J}\,)$
satisfies two general conditions 
in the particle's rest frame. 
First, 
the physical situation should be stationary 
implying the time independence both of
the 4-current and of the fields,
which eliminates the last term in Eq.\ \rf{encons}.
Moreover, 
current conservation simplifies to $\vec{\nabla}\cdot\vec{J}=0$,
so that the most general form of $J^{\mu}(x)$ is given by 
\beq 
J^{\mu}(\vec{r}\hspace{1pt})=\big( \rh(\vec{r}\hspace{1pt}),
\vec{\nabla}\!\times\!\vec{f}(\vec{r}\hspace{1pt})\big),
\label{currentansatz} 
\eeq
where $\rh(\vec{r}\hspace{1pt})$ is  the charge density
and $\vec{f}(\vec{r}\hspace{1pt})$ is an arbitrary vector field.
For example,
the choices $\rh(\vec{r}\hspace{1pt})=q\,\de(\vec{r}\hspace{1pt})$
and $\vec{f}(\vec{r}\hspace{1pt})=\vec{\mu}\:\de(\vec{r}\hspace{1pt})$
describe a point charge $q$ with magnetic moment $\vec{\mu}$
situated at the origin.
Second,
a particle is associated
with the concept of confinement to a small spacetime region,
so that we require $J^{\mu}(\vec{r}\hspace{1pt})$ to be localized
within a finite volume $V_0$.
Outside $V_0$, the 4-current is assumed to vanish rapidly.
Then, 
$\vec{f}(\vec{r}\hspace{1pt})$
can be taken as localized also. 
However, 
other choices for $\vec{f}$ 
are possible 
because adding gradients to $\vec{f}$
leaves $J^{\mu}(\vec{r}\hspace{1pt})$ unaffected.

The zeroth component of Eq.\ \rf{encons}
describes the radiated energy $P^0$.
With our above considerations for $J^{\mu}(\vec{r}\hspace{1pt})$,
the Maxwell equation $\vec{\nabla}\!\times\!\vec{E}=0$,
and the divergence theorem,
this component of Eq.\ \rf{encons} becomes
\beq
\int\limits_{\si} d\vec{\si}\cdot\vec{S}
= -\int\limits_{\si} d\vec{\si}\cdot(\vec{f}\!\times\!\vec{E})\; ,
\label{poy}
\eeq
where we have defined the modified Poynting vector
$\Th_{l0}=S_{l}=-S^{l}$.
Since $\vec{f}(\vec{r}\hspace{1pt})$ is localized,
the right-hand side of Eq.\ \rf{poy} vanishes
for $V$ large enough,
so that energy cannot escape to infinity.
It follows
that the net radiated energy is always zero
in the rest frame of the (prescribed external) charge.
This is unsurprising
because time-translation invariance
is maintained in the rest frame.
The spatial localization of the system
then implies energy conservation.
The $P^0$ flux through any closed surface
must therefore vanish.
As an important consequence of this result,
any nonzero 4-momentum
radiated by a prescribed external charge in uniform motion
is necessarily spacelike.

Next, we investigate the rate $\dot{P}_s=\int_{\si}d\si^{l}\Th_{ls}$
at which 3-momentum is radiated.
The spatial components of Eq.\ \rf{encons}
can be written as
\beq
\:\dot{\!\!\vec{P}}=\int\limits_{V}d^3\vec{r} \;J^{\mu}\vec{\nabla} A_{\mu}\; .
\label{3mom}
\eeq
Here, the necessary manipulations
of the term containing $J^{\mu}$
involve the divergence theorem
and ignoring the resulting surface integral,
which is justified
by the presumed spatial localization.
We continue
by employing
the general solution \rf{Asoln} for $A^{\mu}$
and expressing $J^{\mu}$
in terms of its Fourier expansion 
in the variable $\vec{p}$.
In the limit $V\rightarrow\infty$,
the spatial integration yields
momentum-space delta functions,
and one obtains
\beq
{}\;\;\dot{\!\!\vec{P}}=i\int\limits\fr{d^3 \vec{p}}{(2\pi)^3}
\;\fr{\tilde{J}^{\mu}(-\vec{p}\hspace{1pt})N_{\mu\nu}(\vec{p}\hspace{1pt})\tilde{J}^{\nu}(\vec{p}\hspace{1pt})}
{\vec{p}^{\,4}-4\vec{p}^{\,2}k^2-4(\vec{p}\!\hspace{0.8pt}\cdot\!\vec{k}-i\ve k_0)^2}\;\vec{p}\; ,
\label{restrate}
\eeq
as expected from Parseval's identity.
Inspection shows
that the integrand is odd in $\vec{p}$.
Under the additional condition
that the integrand remains nonsingular for all $\vec{p}$,
the radiation rate vanishes.
This condition is determined
by the denominator of the integrand in Eq.\ \rf{restrate},
which corresponds to the $\om=0$ case
of the dispersion relation \rf{odddisp}.
In a situation involving only timelike wave 4-vectors,
which possess nonzero frequencies in any frame,
regularity of the integrand is ensured
precluding vacuum \v{C}erenkov radiation.
However,
the presence of spacelike wave vectors in our model
typically leads to singularities in the integrand in Eq.\ \rf{restrate}.
Then,
the $i\ve$ prescription
plays a determining role for the value of the integral \rf{restrate}.

A sample charge distribution
yields further insight. 
The general case of a charged magnetic dipole 
is treated in Appendix \ref{md}. 
Here, 
we focus 
on the simpler example of $\vec{J}=\vec{0}$. 
A position-space delta-function source for $\rh$ 
leads to an undesirable asymptotic behavior
of the integrand in Eq.\ \rf{restrate}.
This suggests
to consider a spherical charge $q$
of finite size.
We find the explicit form
\beq
\rh(r)=\fr{q }{4\pi\la^2 r}\exp(-r/\sqrt{2}\la)\sin(r/\sqrt{2}\la)
\label{rho}
\eeq
to be mathematically tractable.
The parameter $\la$ determines the size of the charge.
For a quantum-mechanical particle of mass $m$,
$\la$ might be associated with its Compton wavelength $\la\sim m^{-1}$.
A point particle with delta-function charge distribution
can be recovered in the limit $\la\rightarrow 0$.
The Fourier transform
of $\rh(r)$
is given by $\tilde{\rh}(\vec{p}\hspace{1pt})=q/(\vec{p}^{\, 4}\la^4+1)$.

We perform the integral \rf{restrate}
in spherical-type coordinates
with $\vec{p}=l\;(\sin\th\cos\ph,\sin\th\sin\ph,\cos\th)$.
It is convenient to choose
the polar axis along $\vec{k}$
and to select the integration domain
$l\in [-\infty,\infty]$,
$\th\in[0,\pi/2]$,
and $\ph\in[0,2\pi]$.
The $\ph$ integration is trivial.
The $l$ integral can be evaluated
with the residue theorem.
Closing the contour above or below
yields the same result.
Note
that the two dispersion-relation poles
are shifted in the {\it same} imaginary direction:
depending on the sign of $k_0$,
both poles are either below or above the real-$l$ axis.
The final integration over $\th$
yields the following result:
\beq
\:\dot{\!\!\vec{P}}=-
\fr{{\rm sgn}(k_0)}{16\pi}\;\fr{q^2}{\la^2}\,
{\rm tan}^{-1}(4k_0^2\la^2)\;
\fr{k_0^2}{\vec{k}^{\,2}}\;\vec{e}_k\; ,
\label{samplerate}
\eeq
where $\vec{e}_k$ denotes the unit vector in $\vec{k}$ direction.
One can now take the limit $\la\rightarrow 0$
to obtain the radiation rate for a point charge:
\beq
\:\dot{\!\!\vec{P}}=-{\rm sgn}(k_0)\;
\fr{q^2}{4\pi}\;
\fr{k_0^4}{\vec{k}^{\,2}}\;\vec{e}_k\; .
\label{pcrate}
\eeq
Note
that nonzero rates are possible
despite the antisymmetric integrand in Eq.\ \rf{restrate}.
From a mathematical viewpoint,
this essentially arises
because the physical regularization prescription
(i.e., the contour)
fails to respect this antisymmetry.
The presence of a nonzero flux
in the above static case
may seem counter-intuitive.
However, similar situations in conventional physics
can readily be identified.
For example,
a time-independent situation
with constant non-parallel $\vec{E}$ and $\vec{B}$ fields
is associated
with the nonvanishing Poynting flux $\vec{S}=\vec{E}\!\times\!\vec{B}$.

In the present context,
the absence of the vacuum \v{C}erenkov effect
requires $k_0=0$ in the rest frame.
It follows in particular
that for lightlike $k^{\mu}$,
a point charge never ceases to emit radiation.
This is to be contrasted
with the conventional case,
which typically involves refractive indices
that imply a minimal speed of the charge
for the emission of radiation.
The point is
that vacuum \v{C}erenkov radiation
need not necessarily be a threshold effect.

For many applications,
it is more convenient
to express the radiation rate \rf{pcrate}
in the laboratory frame.
Although the required coordinate change is straightforward,
the resulting expressions
are not particularly transparent
in the general situation.
We therefore focus on
the spacelike-$k^{\mu}$ case
and consider
an inertial frame
in which $k_0'=0$ and $\vec{k}'\neq\vec{0}$
(such a frame always exists).
In what follows,
we can suppress the primes for brevity
because a confusion with the rest-frame components
is excluded.
For finite $\la$, we obtain
\beq
\dot{P}^{\mu}=
\fr{q^2}{16\pi}\;
\fr{\ga(\vec{\be}\!\hspace{0.8pt}\cdot\!\vec{k})^2\la^{-2}}
{\vec{k}^{\,2}+\ga^2(\vec{\be}\!\hspace{0.8pt}\cdot\!\vec{k})^2}\;
{\rm tan}^{-1}\!\big[ 4\la^2\ga^2(\vec{\be}\!\hspace{0.8pt}
\cdot\!\vec{k})^2\big]
\!K^{\mu} ,
\label{labsamplerate}
\eeq
and the point-charge limit gives
\beq
\dot{P}^{\mu}=
\fr{q^2}{4\pi}\;
\fr{\ga^3(\vec{\be}\!\hspace{0.8pt}\cdot\!\vec{k})^4}
{\vec{k}^{\,2}+\ga^2(\vec{\be}\!\hspace{0.8pt}\cdot\!\vec{k})^2}\;
K^{\mu} .
\label{labpcrate}
\eeq
Here, $\vec{\be}$ is the 3-velocity of the charge
in the laboratory
and $\ga$ is the corresponding relativistic gamma factor.
The overdot now denotes laboratory-time differentiation.
The 4-direction
\beq
K^{\mu}\equiv\fr{{\rm sgn}(\vec{\be}\!\hspace{0.8pt}\cdot\!\vec{k})}
{\sqrt{\vec{k}^{\,2}+\ga^2(\vec{\be}\!\hspace{0.8pt}\cdot\!\vec{k})^2}}\;
{ \ga^2(\vec{\be}\!\hspace{0.8pt}\cdot\!\vec{k}) \choose
\;\vec{k}+\ga^2(\vec{\be}\!\hspace{0.8pt}\cdot\!\vec{k})\vec{\be}\; }
\label{dir}
\eeq
arises from transforming
${\rm sgn}(k_0)\vec{e}_k$.
Vacuum \v{C}erenkov radiation is absent
for particle 3-velocities perpendicular to $\vec{k}$.
In all other cases,
both the radiated energy
and the projection of the radiated 3-momentum onto the velocity
are positive,
which decelerates conventional charges
in the chosen laboratory frame.

To determine the polarization of vacuum \v{C}erenkov radiation,
we refer to the discussion of the plane-wave solutions in Appendix \ref{pws}.
In the particle's rest frame,
where the frequency of the emitted waves vanishes,
Eq.\ \rf{EBfields} implies
that the radiated electric field
is purely longitudinal.
In particular,
plane waves emitted along $\vec{k}$
are free of $\vec{E}$ fields,
i.e., they are purely magnetic.
Note
that the energy--momentum tensor \rf{emtensor}
in this case
remains nonzero,
so that such waves are also associated
with a nontrivial momentum flux.

\begin{figure}
\begin{center}
\includegraphics[width=0.8\hsize]{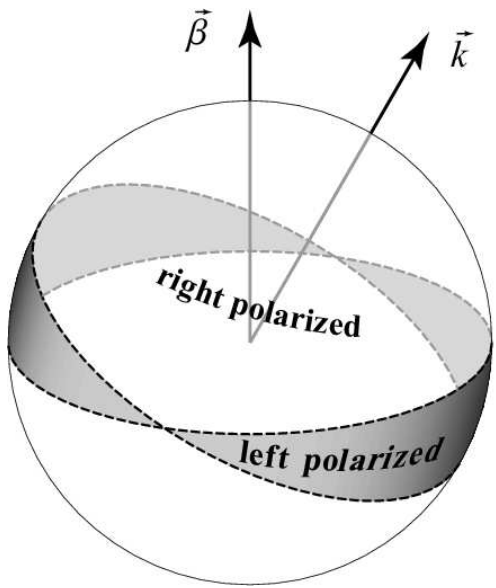}
\end{center}
\caption{Dependence of the polarization on direction.
For vectors $\vec{p}$
pointing in the clear (shaded) direction,
the associated waves are right (left) polarized.
The radiation exhibits linear polarization only
when $\vec{p}$ lies on one of the dashed lines.
Vacuum \v{C}erenkov radiation may not be emitted
into all directions.
The wave 4-vector $p^{\mu}=(\vec{\be}\!\cdot\!\vec{p},\vec{p}\hspace{1pt})$
is further constrained by the dispersion relation \rf{odddisp}.}
\label{ball}
\end{figure}

In our laboratory frame,
where $k^0=0$,
the emitted radiation
is typically left or right polarized,
which can be established as follows.
Consider an emitted wave with a wave vector $\vec{p}$
and select coordinates as in Appendix \ref{pws}.
Equation \rf{EBfields} then yields
\beq
E_y=\fr{p^{\mu}p_{\mu}|\vec{p}\hspace{1pt}|}
{2(\vec{k}\!\hspace{0.8pt}\hspace{0.8pt}\cdot\!\vec{p}\hspace{1pt})
(\vec{\be}\!\cdot\!\vec{p}\hspace{1pt})}iE_x,
\label{pol}
\eeq
where we have used
our result from the previous section
that the 4-vector of a radiated wave
is of the form $p^{\mu}=(\vec{\be}\!\hspace{0.5pt}\cdot\!\vec{p},\vec{p}\hspace{1pt})$.
The direction dependence of the polarization
implied by Eq.\ \rf{pol}
is depicted in Fig.\ \ref{ball}.
We remark
that conventional \v{C}erenkov radiation in isotropic media
is linearly polarized in
the plane spanned by $\vec{\be}$ and $\vec{p}$ \cite{ll}.

The conventional nondispersive \v{C}erenkov case
is associated
with a conical shock wave of opening angle
$\al=\cos^{-1}(c_{ph}/|\vec{\be}|)$.
The present model,
however,
is analogous to a dispersive situation,
where no shock-wave singularity is expected \cite{cerclass,certheo}.
This is supported by the plot in Fig.\ \ref{fig1}.
The concept of a sharply defined \v{C}erenkov cone
is therefore less useful in the present context.

The more interesting question
regarding the magnitude of the wave 3-vector
in a given radiation direction
can be answered as follows.
In general,
the wave vector must satisfy both
the dispersion relation and the \v{C}erenkov condition.
In the laboratory frame,
for example,
the wave 4-vector needs to be of the form
$(\vec{p}\!\hspace{0.8pt}\cdot\!\vec{\be},\vec{p}\hspace{1pt})$
according to the discussion in the previous section.
This yields the constraint equation
\beq
(\vec{p}\!\hspace{0.8pt}\cdot\!\vec{\be}\hspace{1pt})^2=\vec{p}^{\,2}+2\vec{k}^{\,2}
-2\sqrt{(\vec{p}\!\hspace{0.8pt}\cdot\!\vec{k})^2+\vec{k}^{\,4}}
\label{cercone}
\eeq
for a wave 3-vector $\vec{p}$.
Here,
we have again chosen a spacelike $k^{\mu}$
and selected a laboratory frame
with $k^{\mu}=(0,\vec{k})$.
We have further employed the dispersion-relation solution \rf{spacesol},
where the sign choice is restricted
by the \v{C}erenkov condition of spacelike $p^{\mu}$.
For a fixed $|\vec{p}\hspace{1pt}|$,
Eq.\ \rf{cercone} determines a
(distorted)
cone of possible emission directions.
Explicitly, denoting the angle between $\vec\beta$ and $\vec p$ by
$\alpha$ and the angle between $\vec p$ and $\vec k$ by $\theta$,
it follows that
\beq
{\vec{\be}^2\cos^2\alpha-\sin^2\theta\over(\vec{\beta}^2\cos^2\alpha-1)^2}
={|\vec{p}\hspace{1pt}|^2\over 4|\vec{k}\hspace{1pt}|^2}.
\label{cercone2}
\eeq
Note that in
Eq.\ \rf{cercone2}
cylindrical symmetry
about the charge's velocity $\vec{\be}$ is generally lost,
a direct consequence
of rotation breaking due to a nonzero $\vec{k}$.
It is interesting to consider the case 
in which $|\vec{p}\hspace{1pt}|\gg|\vec{k}\hspace{1pt}|$.
Then, Eq.\ \rf{cercone2} gives
\beq
\gamma^{-2}+\sin ^2\alpha
\simeq{2|\vec{\be}\!\cdot\!\vec{k}|\over|\vec{p}\hspace{1pt}|}
\equiv\Lambda^2\ll 1.
\eeq
We conclude that $\gamma>\Lambda^{-1}$,  
which amounts to a minimum speed of the charge
for a given (large) emitted wave 3-vector, 
consistent with the radiation condition in Sec.\ \ref{concept}. 
Note also 
that $\alpha<\Lambda$, 
so that such radiation 
has wave 3-vectors 
within a small cone around $\vec\beta$. 

In a Lorentz-violating situation as well as in conventional dispersive media,
the group velocity and the wave vector
need not be aligned.
It follows
that for a fixed $|\vec{p}\hspace{1pt}|$
the cone determined by Eq.\ \rf{cercone}
does not necessarily coincide with the cone
defined by the motion of the corresponding wave packets.
At a given $|\vec{p}\hspace{1pt}|$,
this
would lead us
to associate the group-velocity cone
rather than the phase-speed cone
with the actual motion of the corresponding emitted disturbance.
However,
such an interpretation would be misleading.
The group velocity
$\vec{v}_g(\vec{p}\hspace{1pt})=\vec{\nabla}_{\!\vec{p}}\, \om$
describes the motion of a wave packet centered at $\vec{p}$ only
when {\it all} momenta $\vec{p}+d\vec{p}$
in the vicinity of $\vec{p}$
contribute to the wave packet.
This is not the case
in the present \v{C}erenkov situation
due to the constraint \rf{cercone}.
In the present case,
the determination of the physical velocity
associated with a disturbance
must only involve momenta $\vec{p}$
that satisfy Eq.\ \rf{cercone}.
This corresponds to the projection of $\vec{v}_g(\vec{p}\hspace{1pt})$
onto the appropriate tangent plane
of the 2-dimensional $\vec{p}\,$-space surface
defined by Eq.\ \rf{cercone}.
This becomes particularly clear
in the charge's rest frame.
Since only waves with $\om=0$ are emitted,
differentiation of $\om$ with respect to a radiated $\vec{p}$
must yield zero.
This is consistent with the time independence of the situation:
any wave pattern must be stationary in the rest frame.

\section{Back-reaction on the charge}
\label{back}

Until now,
the radiating charge $q$ has been considered
as an external prescribed source.
However,
in realistic situations
total energy and momentum are conserved,
so that the radiated 4-momentum
must be supplied by the charged particle,
which will then typically undergo accelerated motion.
This,
in turn,
leads to additional power loss
through the conventional mechanism
described by Larmor's formula \cite{Jackson}.
A more refined analysis
must therefore
include aspects of the particle's dynamics.
Such an analysis
is the topic of the present section.
We simplify the situation
by neglecting acceleration effects,
so that the force acting on the charge
is solely determined
by the back-reaction of vacuum \v{C}erenkov radiation.

To avoid confusion with the wave vector $p^{\mu}$,
we denote the charge's 4-momentum by $Q^{\mu}=(Q^0,\vec{Q})$.
The mass and the 4-velocity of the particle
are  $m$ and $u^{\mu}=\ga(1,\vec{\be}\hspace{1pt})$,
respectively.
We now assume energy--momentum conservation for the \v{C}erenkov
system,
so that
\beq
\dot{Q}^{\mu}=-\dot{P}^{\mu}(\vec{\be}\hspace{1pt}).
\label{ceom}
\eeq
Here,
$\dot{P}^{\mu}(\vec{\be}\hspace{1pt})$ is given by the rate formula \rf{labpcrate}.
For particles with $Q^{\mu}=mu^{\mu}$,
the relation \rf{ceom}
yields their equation of motion
in the form of a first-order differential equation for $\vec{\be}$,
as usual.
However,
in quantum theory,
for example,
the Maxwell--Chern--Simons Lagrangian
may induce a Lorentz-violating dispersion relation
for the charge
through radiative effects.
Then,
momentum and velocity of the particle
need not necessarily be aligned any longer \cite{ck}.
and more care is required.
As a result of our methodology, 
this issue turns out to be nontrivial
even in the present classical context.
Suppose the particle changes its 4-momentum
by $dQ^{\mu}=-\dot{P}^{\mu}dt$
through vacuum \v{C}erenkov radiation.
The method for determining $\dot{P}^{\mu}$ 
discussed in the previous section
then simultaneously fixes \textit{both} 
the change in the charge's energy 
and the corresponding change in its 3-momentum.
We must therefore investigate
the compatibility of our approach 
with the charge's dispersion relation.

To answer this question,
we start with the general momentum--velocity ansatz
$Q^{\mu}=mu^{\mu}+q^{\mu}$,
where $q^{\mu}$ is a Lorentz-violating correction
that can depend on $u^{\mu}$.
Time differentiation,
subsequent contraction with $u_{\mu}$,
and Eq.\ \rf{ceom} yield
$m\dot{u}^{\mu}u_{\mu}+\dot{P}^{\mu}u_{\mu}+\dot{q}^{\mu}u_{\mu}=0$.
Differentiation of $u^{\mu}u_{\mu}=1$
with respect to time establishes
that $\dot{u}^{\mu}u_{\mu}$ is always zero.
In the particle's rest frame,
where the timelike component of $\dot{P}^{\mu}$
and the spacelike components of $u_{\mu}$ vanish,
one verifies
that $\dot{P}^{\mu}u_{\mu}=0$.
We are thus left with
$\dot{q}^{\mu}u_{\mu}=0$
as a constraint for our approach.
Note that this condition
is compatible with
the conventional situation $q^{\mu}=0$,
so that we are allowed to use $Q^{\mu}=mu^{\mu}$.
We remark
that the weakness of this constraint
hinges upon our previous assumption \rf{currentansatz}
of a time-independent current distribution
in the charge's rest frame.
For example,
a rotating-dipole model
of the charge
would lead to energy emission
in the center-of-mass frame,
so that in general $\dot{P}^{\mu}u_{\mu}\neq 0$
requiring a dispersion-relation modification.

We can now proceed using $Q^{\mu}=mu^{\mu}$.
As mentioned above,
this yields the differential equation
\beq
-\dot{P}^{\mu}(\vec{\be}\hspace{1pt})=m\dot{u}^{\mu}(\vec{\be}\hspace{1pt})
\label{eceom}
\eeq
for $\vec{\be}(t)$,
where $\dot{P}^{\mu}(\vec{\be}\hspace{1pt})$
is given by the rate formula \rf{labpcrate}.
It turns out
that this equation can be integrated analytically
in the laboratory frame with $k^{\mu}=(0,\vec{k})$.
The 4-force $-\dot{P}^{\mu}(\vec{\be}\hspace{1pt})$ on the charge
vanishes in the spacelike direction(s) 
orthogonal to $\vec{\be}$ and $\vec{k}$, 
so that the particle's motion 
remains confined 
to the subspace spanned by $\vec{\be}$ and $\vec{k}$. 
The relativistic Newton law \rf{eceom} contains
therefore at most three nontrivial equations:
\bea
-\fr{q^2}{4\pi}\;\vec{k}^{\,2}\fr{\ga^5\be_{\|}^5}{(1+\ga^2\be_{\|}^2)^{3/2}}
& = & m\fr{d}{dt}\ga,\nonumber\\ 
-\fr{q^2}{4\pi}\;\vec{k}^{\,2}\fr{\ga^5\be_{\|}^5}{(1+\ga^2\be_{\|}^2)^{3/2}}
\,\be_{\bot} 
& = & m\fr{d}{dt}\ga\be_{\bot},\nonumber\\ 
-\fr{q^2}{4\pi}\;\vec{k}^{\,2}\fr{\ga^3\be_{\|}^3}{(1+\ga^2\be_{\|}^2)^{1/2}}
\,\be_{\|} 
& = & m\fr{d}{dt}\ga\be_{\|}. 
\label{cceom} 
\eea 
Here, $\be_{\|}$ and $\be_{\bot}$ 
denote the respective magnitudes of the $\vec{\be}$-velocity components 
parallel and perpendicular to $\vec{k}$, 
so that $\ga^{-2}=1-\be_{\|}^2-\be_{\bot}^2$. 

Note
that the three equations of motion \rf{cceom}
determine two unknown functions,
the velocity components 
$\be_{\bot}(t)$ and $\be_{\|}(t)$. 
Compatibility with the dispersion-relation constraint 
discussed above guarantees 
that only two of these equations are independent,
as required by consistency.
To see this explicitly, 
note 
that the first and the second 
of the equations of motion \rf{cceom} 
imply 
\beq 
\be_{\bot}(t)=\be_{\bot}=\text{const.}
\label{perp} 
\eeq
Introducing the variable $\xi\equiv\ga\be_{\|}$
one can now demonstrate
that all three components of the equations of motion \rf{cceom} 
lead to the same differential equation for $\xi$,
and thus $\be_{\|}$, 
given by 
\beq 
-\fr{q^2}{4\pi}\;\vec{k}^{\,2}\sqrt{1-\be_{\bot}^2}\fr{\xi^4}{1+\xi^2}
= m\fr{d}{dt}\xi.
\label{xiode}
\eeq 
This result establishes the dependency among the equations,
and it is suitable for integration.
We obtain 
\beq 
\fr{1}{\xi}+\fr{1}{3\xi^3}=\fr{t+t_0}{\ta}, 
\label{para} 
\eeq 
where $\ta=4\pi m/q^2\vec{k}^{\,2}\sqrt{1-\be_{\bot}^2}$
is the characteristic time scale 
associated with the particle's motion. 
The integration constant $t_0\ge0$ 
is determined by the
velocity of the particle
at $t=0$.
Note that as the time increases, 
the parameter $\xi$, 
and thus $\be_{\|}$, 
decrease, 
so that the charge is always slowed down 
by vacuum \v{C}erenkov radiation
when viewed in our laboratory frame.

Although Eq.\ \rf{para} can be solved 
analytically for $\be_{\|}(t)$, 
the resulting expression is not particularly transparent. 
We therefore consider certain limiting cases. 
Suppose $\xi\gg 1$, 
which corresponds to a fast-moving charge. 
Then, 
the $\xi^{-3}$ term in Eq.\ \rf{para} 
can be neglected and one obtains 
\beq 
\be_{\|}(t)=\fr{4\pi m}{q^2\vec{k}^{\,2}}\;\fr{1}{\sqrt{(t+t_0)^2+\ta^2}}.
\label{lxispeed} 
\eeq 
The corresponding distance $d_{\|}$ traveled parallel to $\vec{k}$ 
is given by 
\beq 
d_{\|}(t)=\fr{4\pi m}{q^2\vec{k}^{\,2}}
\left(\sinh^{-1}\fr{t+t_0}{\ta}-\sinh^{-1}\fr{t_0}{\ta}\right). 
\label{lxidist} 
\eeq 
Since $d_{\bot}=\be_{\bot}t$,
the trajectory is in general no longer a straight line. 
The path of a fast charge 
is determined 
by a hyperbolic-sine function 
with a characteristic scale size of $\be_{\bot}\ta$. 
Such a curved trajectory 
is a direct consequence 
of the involved vacuum anisotropies. 

In the case of a charge moving 
with a nonrelativistic $\be_{\|}$ 
satisfying $\be_{\|}^2\ll 1-\vec{\be}^{\,2}$, 
we have $\xi\ll 1$.  
It follows 
that the $\xi^{-1}$ term in Eq.\ \rf{para} 
is negligible, 
so that 
\beq 
\be_{\|}(t)=\fr{\ta^{1/3}\sqrt{1-\be_{\bot}^2}}
{\sqrt{3^{2/3}(t+t_0)^{2/3}+\ta^{2/3}}}. 
\label{sxispeed} 
\eeq 
This expression can be integrated analytically 
to yield 
\beq 
d_{\|}(t)=\fr{2\pi m}{q^2\vec{k}^{\,2}}
\left[h\left(3\fr{t+t_0}{\ta}\right)-h\left(3\fr{t_0}{\ta}\right)\right] 
\label{sxidist} 
\eeq 
for the distance $d_{\|}$ traveled parallel to $\vec{k}$. 
Here, the function $h$ is given by 
\beq 
h(\chi)\equiv\chi^{1/3}\sqrt{1+\chi^{2/3}}+\sinh^{-1}(\chi^{1/3}).
\label{sxidisth} 
\eeq 
For large $t\gg\ta$, 
the parallel distance traveled increases as $t^{2/3}$,
so that not even asymptotically 
a straight-line trajectory arises. 
If one extrapolates these results 
to the curved-spacetime situation, 
it follows 
that in the Einstein--Maxwell--Chern--Simons system \cite{grav}
a conventional test charge 
would not travel along traditional geodesics 
despite the absence of external electromagnetic fields. 
Such a violation of the equivalence principle 
is seen to be closely tied to the presence of Lorentz breaking.

\section{Phase-space estimate}
\label{phase} 

A quantum-field treatment
of vacuum \v{C}erenkov radiation
would be desirable.
However,
such an analysis
requires a completely satisfactory quantum theory
of the model under consideration,
a condition
that is not met
in most Lorentz-violating frameworks.
In fact,
many approaches to Lorentz breaking 
lack a Lagrangian and are purely kinematical
precluding even a classical analysis
along the lines presented above.
Although such models are theoretically less attractive,
it is still interesting to investigate
to which degree a modified dispersion relation by itself
can give insight into vacuum \v{C}erenkov radiation.

In quantum field theory, 
the rate $\Ga$ for the decay of a particle $P_a$
into two particles,
$P_b$ and $P_c$,
obeys
\beq
d\Ga=\fr{|{\cal M}_{a\rightarrow b,c}|^2}{2E_a}(2\pi)^4
\de^{(4)}(p_a^{\mu}-p_b^{\mu}-p_c^{\mu})d\Pi_bd\Pi_c.
\label{gendecay}
\eeq
Here,
${\cal M}_{a\rightarrow b,c}$ is the transition amplitude
containing information about the dynamics of the decay.
The remaining quantities are associated
with the kinematics of the reaction process.
They include phase-space elements $d\Pi_{s}$
and various 4-momenta $p_s^{\mu}=(E_s,\vec{p}_s)$,
where the subscript $s\in\{a,b,c\}$
refers to the corresponding particle.
In what follows,
we take $P_a$ and $P_b$
to be a charge $q$
with conventional dispersion relation $p_a^2=p_b^2=m^2$.
For comparison 
with the classical result \rf{pcrate}, 
we consider photons $P_c$ 
with a dispersion relation 
corresponding to that of classical plane waves 
in the Maxwell--Chern--Simons model
for lightlike $k^{\mu}$ \rf{lightsol}.
In particular,
we identify the wave frequency $\om$
with the photon energy $E_c$.

To leading approximation,
the determination of ${\cal M}_{a\rightarrow b,c}$
is expected to parallel that of the conventional case:
contraction of the photon polarization 4-vector 
with a $q\ga^{\mu}$-type vertex 
sandwiched between two external-leg spinors.
It follows 
that the amplitude ${\cal M}_{a\rightarrow b,c}$ 
transforms as a coordinate scalar \cite{fn3}. 
As an important consequence, 
the rest-frame and the laboratory-frame amplitudes of a given decay 
cannot differ by relativistic $\ga$ factors, 
which could mask the true energy dependence of the reaction rate. 
Note, however,
that this does not imply 
particle Lorentz symmetry in ${\cal M}_{a\rightarrow b,c}$.
Note also
that ${\cal M}_{a\rightarrow b,c}$ is typically energy dependent:
with our normalization,
the spinor components scale as the square root of their energy,
and the components of photon polarization vectors are of order unity. 
With these considerations, 
we can take ${\cal M}_{a\rightarrow b,c}=qE_aM$
as the generic form of the amplitude \cite{kauf}.
The dimensionless function $M$
is determined by the model's dynamics
and depends on external momenta 
and the Lorentz-violating parameters. 
 
The decay rate $\Ga$, 
defined as the coordinate-scalar transition probability per time,
must pick up a time-dilation factor 
under coordinate boosts. 
On the right-hand side of Eq.\ \rf{gendecay}, 
the $E_a^{-1}$ normalization 
provides this transformation property, 
so that the remaining part of this expression 
is a scalar under observer transformations. 
This implies 
that the phase-space elements $d\Pi_s$ 
must also transform as coordinate scalars 
(see previous footnote \cite{fn3}). 
For Lorentz-symmetric dispersion relations 
and in our normalization, 
$d\Pi$ is determined
by the conventional relation $2E_{\vec{p}}\,(2\pi)^3d\Pi=d^3\vec{p}$,
where $E_{\vec{p}}$ is a dispersion-relation root at $\vec{p}$.
This applies only to the charge in the present example, 
so that $d\Pi_b$ in Eq.\ \rf{gendecay} is conventional. 
However, 
for Lorentz-violating dispersion relations, 
such as that of our photon, 
this expression is no longer coordinate independent. 
It can be verified 
that for the 
positive-energy,
spacelike branches 
of the present modified photon dispersion relation \rf{lightsol} 
the phase-space element 
\beq 
d\Pi_c=\fr{d^3\vec{p}_c}{(2\pi)^32|\vec{p}_c+{\rm sgn}(k^0)\vec{k}|}
\label{psel} 
\eeq
is observer invariant. 
As discussed previously, 
the kinematics of the \v{C}erenkov process
requires spacelike photon 4-momenta, 
so that Eq.\ \rf{psel} 
is indeed the relevant one in the present context. 
Note 
that for zero $k^{\mu}$ 
the conventional expression is recovered. 

With our above considerations
Eq.\ \rf{gendecay} becomes
\beq 
d\Ga=\fr{q^2|M|^2m}{32\pi^2}
\;\fr{\de^{(4)}(p_a^{\mu}-p_b^{\mu}-p_c^{\mu})\:d^3\vec{p}_b\:d^3\vec{p}_c}
{\sqrt{\vec{p}_b^{\,2}+m^2}\;|\vec{p}_c+{\rm sgn}(k^0)\vec{k}|}
\label{restdecay}
\eeq
in the charge's rest frame. 
The $\vec{p}_b$ integration 
can be performed straightforwardly. 
For the $\vec{p}_c$ integral 
we select spherical coordinates 
with $\vec{k}$ along the polar axis.
We denote the azimuthal and polar angles 
by $\th$ and $\ph$, 
respectively. 
The limit of a nondynamical charge, 
which is necessary for comparison 
with the previous classical treatment, 
can be recovered here for $m\to\infty$.
Then, 
the remaining delta function
gives the constraint
$|\vec{p}_c|=-2\:k^0\cos\th$. 
We remark  
that this condition implies zero-energy photons 
as decay products 
consistent with our previous dynamical analysis 
in the classical context. 
Note also 
that this constraint restricts the angular integrations 
to the upper or lower hemisphere depending on the sign of $k^0$. 
The $|\vec{p}_c|$ integration
now yields
\beq 
\fr{d\Ga}{d\Om}=-\fr{q^2|M|^2}{8\pi^2}k^0\cos\th
\label{ddrate} 
\eeq
for the differential decay rate 
in the charge's rest frame. 
Here, $\Om$ denotes the solid angle.

For the remaining angular integrations,
the functional dependence $M=M(\Om)$ is needed. 
However, 
we are interested 
in a rough approximation for the decay rate only, 
so that it appears reasonable 
to replace $|M(\Om)|^2$
by a constant $\overline{|M|^2}\sim{\cal O}(1)$ 
corresponding perhaps 
to a suitable angular average of $|M(\Om)|^2$. 
Then, 
an estimate for the total decay rate is given by 
$\Ga\simeq(8\pi)^{-1}\overline{|M|^2}q^2|\vec{k}|$. 
Noting  
that $d\dot{P}^{\mu}=p^{\mu}(\Om)\,d\Ga$,
the rate of momentum emission 
can be determined similarly. 
We obtain
\beq 
\:\dot{\!\!\vec{P}}\simeq-{\rm sgn}(k^0)
\fr{q^2\overline{|M|^2}}{8\pi}\vec{k}^{\,2}\vec{e}_k
\label{momdecay} 
\eeq
as an estimate for the net radiated momentum per time 
in the charge's rest frame. 
Comparison with Eq.\ \rf{pcrate} 
obtained from the corresponding classical treatment reveals 
agreement 
(up to the indeterminate numerical factor $\overline{|M|^2}$) 
with the above phase-space estimate \rf{momdecay}. 

This result demonstrates 
that a careful kinematical analysis 
of the \v{C}erenkov decay 
can give a sensible estimate 
for the rate of momentum emission. 
Note, 
however, 
the assumptions involved: 
equality of plane-wave and one-particle dispersion relations, 
absence of additional symmetries suppressing the quantum amplitude, 
and a nondynamical Lorentz-symmetric charge $m\rightarrow\infty$.
We also emphasize 
the importance of constructing invariant phase-space elements 
in the present context. 

\section{Experimental outlook}
\label{exp}

This section mentions some examples
of potentially observable signatures
for Lorentz violation
in the context of vacuum \v{C}erenkov radiation.
Such experimental effects
can be grouped into two broad classes:
detection of the emitted radiation itself 
through its properties analyzed in Sec.\ \ref{rate} 
and effects on the charge's motion
discussed in Sec.\ \ref{back}.

Searches for the emitted electromagnetic radiation 
are perhaps suggested 
by the analogous discovery 
of the conventional \v{C}erenkov effect. 
In the present case, 
a fast charged particle 
should radiate left or right polarized waves 
into directions determined by Eq.\ \rf{cercone}. 
The phase-speed condition 
and the dispersion relation (\ref{odddisp})
imply that
the maximum frequency
emitted obeys
$\om_{\textrm{max}}\lsim\ga^2\:{\cal O} (k^{\mu})$. 
Here, $\ga$ is the boost factor 
corresponding to the charge's speed 
and ${\cal O} (k^{\mu})$ 
denotes the typical size of $k^{\mu}$ components 
in the laboratory frame, 
which are observationally constrained by 
${\cal O} (k^{\mu})\lsim 10^{-42}\,$GeV \cite{mcsclass}. 
Taking this bound to be saturated, 
we find for the example a proton 
at the end of the observed cosmic-ray spectrum ($10^{20}\,$eV) 
that $\om_{\textrm{max}}$ 
is of the order of $1.6\times 10^{4}\,$rad/s
corresponding to a wavelength of $1.2\times 10^{5}\,$m.
On a speculative note, 
such radiation might perhaps be observable 
in high-energy astrophysical jets 
emitted in the direction of sight. 

The presence of Lorentz violation
in electrodynamics
can also affect the motion of particles
via the vacuum \v{C}erenkov effect.
For example,
a high-energy charge
would be slowed down
due to the emission of radiation.
This would lead to an effective cut-off
in the cosmic-ray spectrum
for primary particles carrying an electric charge
or a magnetic dipole moment.
This idea has been widely employed in the literature
to place bounds on Lorentz breaking.
In the present Maxwell--Chern--Simons model,
which is already tightly constrained by other considerations,
the energy-loss rate
is suppressed by two powers
of the Lorentz-violating coefficient $k^{\mu}$.
However,
it would be interesting
to consider the dimensionless $k_{F}$ term in the SME:
some of its components 
are currently only bounded at the $10^{-9}$ level \cite{cavexpt},
and a dynamical study paralleling the present one
could yield less suppressed rates.

Another potential signature
associated with the charge's motion
is of statistical nature.
Consider, 
for instance, 
Eq.\ \rf{labpcrate},
which is valid in a laboratory frame
with purely spacelike $k^{\mu}=(0,\vec{k})$.
In this frame,
particles with velocities $\vec{\be}$
that are perpendicular to $\vec{k}$
cease to radiate.
Thus,
the presence of a spacelike $k^{\mu}$
during most of the cosmological history
would constrain the average motion of charges
to 3-velocities lying in a two-dimensional plane.
This effect might be more efficient
before electroweak symmetry breaking
for two reasons.
First,
radiation is not yet decoupled from the matter,
so that there are a large number of free charges
that can be affected.
Second, massless charged matter
is associated with lightlike 4-momenta
so that all wave frequencies
can contribute to vacuum \v{C}erenkov radiation.
Investigations in such a context
might therefore provide
stringent complementary Lorentz-violation bounds.

\section{Conclusions}
\label{conc}

Lorentz-violating vacua
can arise in various approaches to fundamental physics
and are described at low energies by the SME.
In this paper,
we have considered
the physics of electrodynamics in such vacua,
which exhibits close parallels
to the conventional Maxwell case
in macroscopic media.
Our study has focused on vacuum \v{C}erenkov radiation,
which is the analogue of the usual \v{C}erenkov effect
whereby light is emitted
from charges moving uniformly
with superluminal speeds in a medium.
Although we have performed our analysis
primarily within the classical Maxwell--Chern--Simons limit of the SME,
we expect most results and our methodology
to remain applicable
in more general cases
including non-electromagnetic ones.

In Sec.\ \ref{concept},
we have developed a qualitative physical picture
of the \v{C}erenkov effect for general situations
that augments the usual one in macroscopic media.
It also permits
an alternative extraction the general radiation condition:
energy--momentum transport to infinity,
and thus radiation,
can only occur
when the fields fall of like $r^{-1}$.
This in turn requires
that purely spacelike wave 4-vectors
satisfy the electromagnetic plane-wave dispersion relation
in the charge's rest frame.

Based on this intuitive physical picture,
we have developed a method
complementing the conventional one
for the determination of the 4-momentum flux
associated with \v{C}erenkov radiation.
The advantage of this procedure lies in the fact
that it does not require the explicit knowledge of the far fields.
In the Maxwell--Chern--Simons model,
our method permits the calculation
of the exact 4-momentum radiation rate
for a prescribed point charge with magnetic moment.
The corresponding polarization
of the emitted vacuum \v{C}erenkov radiation
follows from Eq.\ \rf{pol}
and is depicted in Fig.\ \ref{ball}.

Section \ref{back}
has treated the effects
of vacuum \v{C}erenkov radiation
on the charge.
Within our framework,
the back-reaction
provides a nontrivial constraint
on the dispersion relation of the charge.
For reasonable models of the charge's
4-current distribution,
this constraint is mild enough
to allow the conventional Lorentz-symmetric
3-momentum dependence of the energy.
As part of our discussion,
we have determined the modified
trajectory of a point charge
in the Maxwell--Chern--Simons model,
which remains no longer a geodesic.
This type of equivalence-principle violation
appears generic
in the presence of vacuum \v{C}erenkov radiation.

In some situations,
useful insight into the \v{C}erenkov-radiation rate
can be obtained
by purely kinematical considerations.
We have exemplified this in Sec.\ \ref{phase}
by a detailed phase-space estimate
that involves photons
obeying the Maxwell--Chern--Simons
plane-wave dispersion relation.
We found
that the construction of coordinate-independent phase-space elements
in the presence of Lorentz breaking
is an important nontrivial issue.
when Lorentz symmetry is violated.

\acknowledgments
We thank D.\ Sudarsky for discussion.
This work was supported in part
by the Centro Multidisciplinar de Astrof\'{\i}sica (CENTRA)
and by the Funda\c{c}\~ao para a Ci\^encia e a Tecnologia (Portugal)
under Grant No. POCTI/FNU/49529/2002.

\appendix

\section{Dispersion relation}
\label{dr}

For a given wave 3-vector $\vec{p}$,
the dispersion relation \rf{odddisp}
determines the corresponding values for $\om$.
In the lightlike-$k^{\mu}$ case,
one obtains
\beq
\om^{(n)}_{\pm}=\pm|\vec{p}+(-1)^{n}\vec{k}|-(-1)^{n}k^0,
\label{lightsol}
\eeq
where $n\in\{1,2\}$.
For purely spacelike $k^{\mu}=(0,\vec{k})$,
the plane-wave frequencies are given by
\beq
\om^{(n)}_{\pm}=\pm\sqrt{\vec{p}^{\,2}+2\vec{k}^{\,2}
+2(-1)^{n}\sqrt{\vec{k}^{\,4}+(\vec{p}\!\hspace{0.8pt}\cdot\!\vec{k})^2}}.
\label{spacesol}
\eeq
If $k^{\mu}=(k^0,\vec{0})$ is purely timelike,
the solutions of the dispersion relation are
\beq
\om^{(n)}_{\pm}=\pm\sqrt{\vec{p}^{\,2}+2(-1)^{n}k^0|\vec{p}\hspace{1pt}|}.
\label{timesol}
\eeq
Note
that branches determining spacelike wave 4-vectors
occur in each of the canonical cases
\rf{lightsol}, \rf{spacesol}, and \rf{timesol}.

For general $k^{\mu}$,
the exact roots of the dispersion relation \rf{odddisp}
can also be obtained straightforwardly.
However,
they are less transparent,
so that we only give expressions
that are correct to first order 
in the Lorentz-violating parameter $k^{\mu}$.
Without loss of generality
we can rotate the coordinate system
such that $p^{\mu}=(\om,0,0,p_z)$
and $k^{\mu}=(k^0,|\vec{k}|\sin\th,0,|\vec{k}|\cos\th)$.
The approximate solutions are then given by
\beq
\om^{(n)}_{\pm}=\pm p_z+(-1)^{n}(k^0\mp|\vec{k}|\cos\th).
\label{gensol}
\eeq
We note that
we have selected convenient labels for these first-order results,
which do not necessarily correspond to the
labels for the canonical cases discussed earlier.

\begin{figure}
\begin{center}
\includegraphics[width=0.8\hsize]{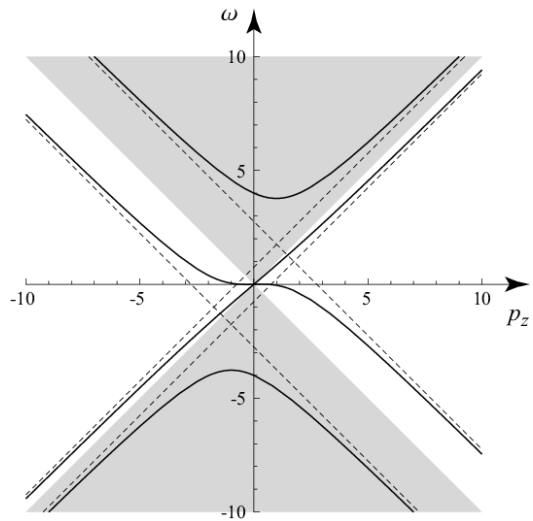}
\end{center}
\caption{Sample solution of the plane-wave dispersion relation.
The solid lines correspond to the exact roots.
The first-order solutions are shown as broken lines.
The shaded region represents
the interior of the $p^{\mu}$-space lightcone.
}
\label{gendr}
\end{figure}

In Fig.\ \ref{gendr},
the plane-wave frequencies $\om^{(n)}_{\pm}$
are plotted versus $p_z$ in some appropriate units
for $k^0=1$, $|\vec{k}|=2$, and $\th=1/2$.
The solid lines
represent the four branches of the exact roots
and the broken ones
the corresponding first-order solutions \rf{gensol}.
The interior of the $p^{\mu}$-space lightcone
has been shaded.
For a given nonzero wave 3-vector,
there are two timelike and two spacelike wave 4-vectors
satisfying the dispersion relation,
as expected from the above discussion
of the three canonical cases.

\section{Plane-wave solutions}
\label{pws}

A plane-wave ansatz $A^{\nu}(x)=A_p^{\nu}(p)\exp(-ip\!\hspace{0.8pt}\cdot\!x)$
in Eq.\ \rf{oddeom}
for $j^{\mu}=0$ gives
\beq
\left(p_{\mu} p_{\nu}-p^2 \et_{\mu\nu}
+2i\ve_{\mu\nu\rh\si}k^{\rh}p^{\si}\right)A_p^{\nu}(p)
=0.
\label{feqom}
\eeq
In what follows,
we adopt Lorentz gauge $p\!\cdot\!A=0$
and the following coordinates:
$\vec{p}=p_z\vec{e}_z$ and $\vec{k}=k_x\vec{e}_x+k_z\vec{e}_z$,
where $\vec{e}_x$ and $\vec{e}_z$ are the usual unit vectors
in the 1- and 3-direction, respectively.
For $p^{\mu}$ satisfying the plane-wave dispersion relation \rf{odddisp},
the polarization vectors
\beq
A_p^{\nu}(\om,\vec{p}\hspace{1pt})=\fr{g}{2p_z}
\left(\begin{array}{c}
2k_xp_z\\ 2(k_0p_z-k_z\om)\\ -i(\om^2-p_z^2)\\ 2k_x\om
\end{array}\right)
\label{Avec}
\eeq
obey Eq.\ \rf{feqom}, where $g$ is a constant.

The electric field
$\vec{E}(x)=\vec{E}_p(\om,\vec{p}\hspace{1pt})\exp(-ip\!\hspace{0.8pt}\cdot\!x)$
and the magnetic field
$\vec{B}(x)=\vec{B}_p(\om,\vec{p}\hspace{1pt})\exp(-ip\!\hspace{0.8pt}\cdot\!x)$
of the plane wave
are now determined
by the conventional field--potential relationship,
so that the polarization vectors are
\bea
\vec{E}_p(\om,\vec{p}\hspace{1pt}) & = & \fr{ig}{2p_z}
\left(\begin{array}{c}
2\om(k_0p_z-k_z\om)\\ -i\om(\om^2-p_z^2)\\ 2k_x(\om^2-p_z^2)
\end{array}\right),\nonumber\\
\vec{B}_p(\om,\vec{p}\hspace{1pt}) & = & \fr{ig}{2}
\left(\begin{array}{c}
i(\om^2-p_z^2)\\ 2(k_0p_z-k_z\om)\\ 0
\end{array}\right).
\label{EBfields}
\eea
The physical fields are understood
to be given by the real parts 
of the resulting plane-wave expressions, 
as usual.
The magnetic field
remains transverse 
because the homogeneous equation $\vec{\nabla}\!\cdot\!\vec{B}=0$
is unaltered.
Note,
however,
that the electric field can exhibit
longitudinal components.

In conventional optics,
a plane wave is called left (right) polarized,
when the electric-field vector
rotates (counter)clockwise
around the wave vector $\vec{p}$
at a fixed point in space
for an observer looking in the direction of propagation \cite{Jackson}.
In the present context,
we adopt the analogous definition
involving the motion
of the transverse electric-field component $\vec{E}_{\bot}$.
One can then distinguish
between elliptical polarization,
and the limiting cases
of linear and circular polarization,
as usual.
An important example
is the case
in which the wave vector $\vec{p}$
is large compared to the components of $k^{\mu}$.
Then, Eqs.\ \rf{gensol} and \rf{EBfields} give
$E_y=\mp(-1)^{n} iE_x$,
where the upper (lower) sign
corresponds to waves of positive (negative) frequency.
The longitudinal component of $\vec{E}$ vanishes in this limit.
It follows
that such waves exhibit the conventional circular polarizations.

Note
that the above definition of polarization
can fail in certain circumstances.
For instance,
it follows from Eq.\ \rf{EBfields}
that zero-frequency waves,
such as \v{C}erenkov radiation in the charge's rest frame,
are associated with $\vec{E}_{\bot}=\vec{0}$.
The electric field is then purely longitudinal
(or zero)
precluding any of the transverse polarizations.
Although it leaves unaffected the polarization,
we also remark that for waves with a phase speed $c_{ph}<1$
the direction of propagation,
which is involved in the polarization definition,
is observer dependent.

\vspace{20pt}
\section{Charged magnetic dipoles}
\label{md}

In this appendix,
we refine our model of the charged particle
by including a magnetic moment $\vec{\mu}$
into our analysis.
This is phenomenologically interesting
because all known electrically charged elementary particles
carry nonzero spin,
which is associated with a finite magnetic moment.
Moreover,
the $q=0$ limit of the model
then describes the \v{C}erenkov effect
in the presence of neutral particles
with magnetic moments,
such as neutrons.

The rest-frame current distribution $J^{\mu}$
of a point magnetic dipole $\vec{\mu}$
with charge $q$
located at the origin
is given by
$J^{\mu}(\vec{r}\hspace{1pt})=\big(q\,\de(\vec{r}\hspace{1pt}),
-\vec{\mu}\!\hspace{1.2pt}\times\!\vec{\nabla}\de(\vec{r}\hspace{1pt})\big)$,
as mentioned previously in the discussion of ansatz \rf{currentansatz}.
To force convergence in certain intermediate steps of the calculation,
we write $\de(\vec{r}\hspace{1pt})=\lim_{\la\to 0}f(r,\la)$
for the delta function,
where
\beq
f(r,\la)=\fr{1}{4\pi\la^2 r}
\exp(-r/\sqrt{2}\la)\sin(r/\sqrt{2}\la)
\label{reg}
\eeq
paralleling the pure-charge case in Sec.\ \ref{rate}.
We remark
that the magnetic-moment definition
$\vec{\mu}=\half\int\vec{r}\!\hspace{0.5pt}\times\!\vec{J}\:d^3r$
in classical electrodynamics \cite{Jackson}
is consistent
with our choice of current
for all $\la$.
The charge density,
and thus its Fourier image,
remain unchanged
relative to those used in Sec.\ \ref{rate}.
The 3-current $\vec{J}(\vec{r}\hspace{1pt})$
takes the form
${}\,\tilde{\!\vec{J}}(\vec{p}\hspace{1pt})=i(\vec{p}\!\hspace{0.8pt}\times\!\vec{\mu})
(\la^4\vec{p}^{\,4}+1)^{-1}$
in Fourier space.

Next,
we use Eq.\ \rf{restrate}
to obtain an explicit integral expression
for the radiation rate in the dipole's rest frame:
\begin{widetext}
\beq
\:\dot{\!\!\vec{P}}=i\int \fr{d^3 \vec{p}}{(2\pi)^3}
\;\fr{\vec{p}^{\,2}\big[q^2-(\vec{p}\!\hspace{0.8pt}\times\!\vec{\mu})^2\big]
+4q\big[(\vec{k}\!\hspace{0.8pt}\cdot\!\vec{p}\hspace{1pt})
(\vec{\mu}\!\hspace{0.8pt}\cdot\!\vec{p}\hspace{1pt})
-(\vec{k}\!\hspace{0.8pt}\cdot\!\vec{\mu})\vec{p}^{\,2}\big]
-4\big[\vec{p}\!\hspace{0.8pt}\cdot\!(\vec{k}\!\hspace{0.8pt}\times\!\vec{\mu})\big]^2
-4q^2k_0^2} 
{\big[\vec{p}^{\,4}-4\vec{p}^{\,2}k^2
-4(\vec{k}\!\hspace{0.5pt}\cdot\!\vec{p}-i\ve k_0)^2\big]
\big[\la^4\vec{p}^{\,4}+1\big]^2}\;\vec{p}\; .
\label{dipolerateexp} 
\eeq 
\end{widetext}
We perform this integration 
in spherical-type coordinates
with $\vec{p}=l\;(\sin\th\cos\ph,\sin\th\sin\ph,\cos\th)$.
We select the polar axis along $\vec{k}$,
and $\vec{\mu}$ lies in the $xz$ plane 
such that $\vec{\mu}=|\vec{\mu}\hspace{.5pt}|\,(\sin\al,0,\cos\al)$.
To apply complex-integration methods, 
we chose the integration domain 
$l\in [-\infty,\infty]$,
$\th\in[0,\pi/2]$, 
and $\ph\in[0,2\pi]$, 
as before. 
The $l$ integral can then be evaluated
with the residue theorem. 

As discussed before,
finite emission rates
can arise only
in the presence of poles at real $l$.
Only the dispersion-relation part
of the denominator in the integral \rf{dipolerateexp}
with zeros at $l_{\pm}=\pm 2(k_0^2-\vec{k}^{\,2}\sin^2\th)^{1/2}$
can lead to such poles.
The corresponding values for $\th$
that also lie within the above range of integration
are determined by $(1-k_0^2/\vec{k}^{\,2})^{1/2}\le\cos\th\le 1$.
In this case,
the contour for the $l$ integration
is fixed
by the causal $i\ve$ prescription:
up to an unimportant normalization of $\ve$,
the poles are shifted to
$l_{\pm}\to\, l_{\pm}-i\ve\,\text{sgn}(k_0)$.
Suppose $k_0>0$,
so that the contour passes above the real poles at $l_{\pm}$.
We then choose to close the integration contour above
encircling the poles at $l^{\pm}=(\pm 1 + i)/\sqrt{2}\la$
with the respective residues $R^{\pm}(\la,\al,\th,\ph)$.
It is now straightforward 
to evaluate the $l$ integral 
with the aid of the residue theorem.
We obtain
\beq
{}\;\;\dot{\!\!\vec{P}}=-
\!\!\!\!\!\!\!\!\!\!\!\int\limits_{\sqrt{1-k_0^2/\vec{k}^{\,2}}}^{1}
\!\!\!\!\!\!\!\!\!\!d\cos\th
\!\int\limits_{0}^{2\pi}\!d\ph\;
\fr{R^{+}+R^{-}}{(2\pi)^{2}}
\left(\begin{array}{c}\sin\th\cos\ph\\
\sin\th\sin\ph\\
\cos\th
\end{array}
\right).
\label{pint}
\eeq
We remark
that an analogous calculation in situations with $k_0<0$
gives the same expression with the opposite sign
provided the symmetries of the residues are taken into account.

For further progress, 
we use the explicit form of the residues 
$R^{\pm}(\la,\al,\th,\ph)$ 
and take the point-particle limit $\la\to 0$. 
This permits a closed-form 
evaluation of the remaining angular integrals: 
\begin{widetext}
\beq 
{}\;\dot{\!\!\vec{P}}=-\fr{{\rm sgn}(k_0)}{12\pi}\; 
\fr{k_0^5}{|\vec{k}|^5}
\left\{
\left[3q^2 \vec{k}^{\,2}/k_0^2 +6q\,\vec{k}\!\hspace{0.8pt}\cdot\!\vec{\mu}
-\vec{\mu}^{\,2}k_0^2
+5(\vec{k}\!\hspace{0.8pt}\cdot\!\vec{\mu})^2k_0^2/\vec{k}^{\,2}
+10(\vec{k}\!\hspace{0.8pt}\times\!\vec{\mu})^2\right]k_0\vec{k}
-2\left[q\vec{k}^{\,2}/k_0^2+\vec{k}\!\hspace{0.8pt}\cdot\!\vec{\mu}\right]k_0^3\,\vec{\mu}\right\}.
\label{drate} 
\eeq 
\end{widetext}
Equation \rf{drate} 
gives the exact expression 
for the net momentum 
radiated by a charged pointlike magnetic dipole 
as measured in its rest frame. 
In the $\vec{\mu}\to\vec{0}$ limit, 
our previous result \rf{pcrate} 
for a point charge is recovered. 
The leading-order corrections
to the point-charge rate \rf{pcrate}
arising from the presence of the magnetic moment $\vec{\mu}$
are suppressed
by an additional power of $k^{\mu}$,
as expected on dimensional grounds.
Although heavily suppressed by four powers of $k^{\mu}$,
the rate does remain nonzero
in the pure-dipole limit $q\to 0$.


\begin{thebibliography}{xx}
\bibitem{fn1} 
We remark, 
however, 
that in unrealistic situations with spacetime constant $j^{\mu}$
the right-hand side of Eq.\ \rf{emcurrent} 
is given by the total divergence 
$-\prt_{\mu}(\et^{\mu\nu}j\!\cdot\!A-j^{\mu}A^{\nu})$. 
The expression in parentheses 
can then be identified 
with the energy--momentum tensor 
associated with the field--source interaction. 
In this case,
the total 4-momentum
stored both in
the fields and in the interaction is conserved,
as expected from translational symmetry.
\end{thebibliography}
\end{document}